\documentclass[a4paper,article]{mdpi1}
\usepackage{mydefs}

\Title{Regional Coverage Analysis of LEO Satellites with Kepler Orbits}

\Author{Iestyn Morgan-Jones$^1$ and Pavel Loskot$^1$}

\address{$^{1}$ College of Engineering, Swansea University, Swansea, UK}

\abstract{Low Earth orbit (LEO) satellites are being considered for expanding
  legacy terrestrial cellular networks. The end users may not be able to
  optimize satellite orbits and constellations, however, they can optimize
  locations of ground stations which aggregate terrestrial traffic and
  inter-connect with over-passing satellites. Such optimization requires a
  model of satellite visibility to decide when the satellite becomes visible to
  the ground station in a given geographical location. Our model assumes ideal
  Kepler orbits parameterized by six orbital elements to describe the satellite
  movements. The steps of Kepler orbit modeling are presented in detail to
  enable other studies requiring geometric representation of satellite
  trajectories in the 3D space or their 2D footprint on the Earth surface. The
  long-term mean time satellite visibility (MTSV) metric is defined to measure
  satellite visibility at a given ground station. Numerical results reveal that
  efficiency of optimizing the ground station locations is dependent on the
  satellite orbit and other satellite visibility constraints. The ground
  station location optimization is especially important when MTSV is improved
  by orthogonal time sharing schemes using multiple satellites on the same or
  different orbits. Similar conclusions can be drawn assuming other performance
  metrics such as the capacity of links between the ground station and the
  satellites.}


\begin{document}
\highlights{
  \begin{itemize}
  \item A detailed model of Kepler orbit is presented, so it can be implemented
    in commonly available simulation software.\smallskip
  \item Kepler orbit model yields celestial trajectories of satellites in the
    sky as well as the ground footprint trajectories.\smallskip
  \item Simulating satellite trajectories is important to design and analyze
    satellite communication networks, to understand spatial geometry of
    satellite constellations, to model more realistic satellite communication
    channels and the associated terrestrial coverage, and to investigate
    communication protocols in satellite networks with given topology.
  \end{itemize}
}

\abbreviations{
  \indent
  \begin{tabular}{@{}ll}
    ECEF & Earth-centered Earth-fixed coordinates \\
    ECI & Earth-centered inertial coordinates \\
    GHA & Greenwich hour angle \\
    GMST & Greenwich mean sidereal time \\
    LEO & low Earth orbit \\
    MRE & minimum required elevation \\
    MTSV & mean time satellite visibility \\
    RAAN & right ascension of ascending node \\
  \end{tabular}
}

\constants{
  \begin{tabular}{@{}ccl}
    $398600.4$ & $\units{km^3/s^2}$ & standard gravitational parameter
    \cite{wkp:standardgravpar}\\
    $\num{6.3781e6}$ & $\units{m}$ & Earth radius \cite{wkp:earthrad}\\
    $\num{7.2921150e-5}$ & $\units{rad/s}$ & Earth rotation angular speed
    \cite{wkp:earthrot} \\
    $\num{1.7275643}$ & $\units{rad}$ & angular offset for Greenwich mean
    sidereal time \cite{wkp:sidertime}\\
  \end{tabular}
}\vspace{24pt}

\noindent\rule{\textwidth}{0.6pt}

%%%
\section{Introduction}
%%%

Satellite technology is getting progressively cheaper due to advances in making
the satellites smaller and lighter. This trend is supported by new aerospace
enterprises who are starting to offer the cost affordable satellite launches
into the orbit. Telecommunication services enabled by LEO satellite networks
are particularly attractive due to their relative proximity to the Earth
surface which reduces signal attenuations. The key design considerations for
satellite networks are choosing appropriate satellite constellation,
maintaining stable orbits, dynamically allocating frequency channels, and
managing the connectivity and interference to terrestrial stations as well as
in between satellites. However, satellite communication services are still
rather expensive due to high initial (CapEx) and operational expenditures
(OpEx), and the time constrained bandwidth sharing by a large pool of
terrestrial users. In order to reduce the deployment times and costs while also
minimizing the need for propulsive corrections at the orbit, small scale
satellites are often launched together in batches and separated at the orbit to
form the desired constellation \cite{crisp2015}.

The vision of a dense satellite network to provide the ubiquitous Internet
access all over the Earth including basic capacity calculations for such
network is presented in \cite{khan16}. The Quality-of-Service (QoS) issues in
satellite mesh networks with delay-sensitive as well as delay-tolerant traffic
are studied in \cite{lee2019} assuming the link layer and network layer
protocols. The link state routing for satellite networks having regular
topology and embracing the whole Earth while experiencing uneven traffic
distribution over different geographical areas is investigated in
\cite{wang2019}. However, none of these works assume specific satellite
trajectories. Precise observations of satellite trajectories require detailed
understanding of orbital mechanics \cite{satobs}. The main but not all steps
for calculating Kepler orbits are presented in \cite{zantou2005}. Comprehensive
review of satellite orbital mechanics is provided, for example, in
\cite{chobotov02} and in \cite{vallado1997}. At lower altitudes up to about 800
km above the Earth surface, satellite orbits are more complex as their
trajectories are subject to orbital decays due to residual atmospheric drag
\cite{bowman2003}. The upper atmosphere and gravity field are causing small
changes in satellite movements referred to as perturbations. The Lagrangian
perturbation forces are studied in \cite{gergely2015} to generalize Kepler laws
of idealistic satellite motions to more realistic scenarios. The description of
satellite trajectories, their ground tracks, orbits relative to the Earth, and
modeling Earth views from the satellites can be found in \cite{capderou05}.

The problem of satellite constellation design to achieve the full Earth
coverage has been addressed in \cite{ballard80}. The geometries of satellite
constellations as a function of the inter-satellite distances and the maximum
latitude to provide the continuous Earth coverage are investigated in
\cite{danesf2006}. The capacity of satellite communication network created by
some satellite constellations in different orbits is evaluated in
\cite{werner1995}. The satellite constellation to maximize visibility from a
given group of ground terminals is optimized in \cite{chen17}. The methodology
for calculating the mean visibility of satellite network from a ground station
possibly equipped with the tracking antenna is developed in \cite{itu97}. The
long-term optimum locations of terrestrial stations having optical
communication links to multiple satellites are investigated in
\cite{lyras2018}. The optical satellite channels are affected by the presence
of atmospheric clouds, so switching among terrestrial stations creates the
transmission diversity. Accurate approximation of the closest approach for LEO
satellites with the largest elevation to the ground terminal is derived in
\cite{ali99}. However, neither other orbital in-view periods which may also
provide connectivity nor the minimum required elevation of the satellite above
the horizon are considered. Typical value of the minimum elevation for a
satellite to become visible is about $10^{\circ}$ \cite{ali1998, danesf2006}.
The coverage of a single LEO satellite at the fixed sky location is obtained in
\cite{pratt99}, and later elaborated in \cite{cakaj14}. Assuming the circular
LEO satellite orbits with uniformly distributed ground stations, the
distribution of satellite in-view times are obtained in \cite{seyedi12}.
However, the analyses in \cite{pratt99, cakaj14,seyedi12} are limited in
accounting for generally spatially non-synchronous trajectories of LEO
satellites where every additional satellite overpass over the ground station is
further shifted to the East or to the West.

Since satellites move over complex orbits, maintaining stable inter-satellite
connectivity in large satellite constellations is rather challenging
\cite{hossain2017}. The link performance in satellite channels is dominated by
shadowing. Using empirical measurements, it was shown in \cite{lutz1991} that
satellite links fluctuate between good and bad transmission channels. The
scattered multi-path components in satellite channels can be assumed to be
Rayleigh distributed whereas Nakagami distribution is a good fit to empirical
measurements for the line-of-sight component \cite{abdi2003}. The Doppler
frequency shift of satellite channels is proportional to the satellite
elevation above the ground station \cite{ali1998}.

Even though satellite constellations and orbits can be designed to maximize
their visibility over a given geographical region, such option is rarely
available to majority of terrestrial users. The users can, however, optimize
the location of ground observation stations. Hence, in this paper, satellite
visibility over a whole geographical region is evaluated numerically, so the
locations of terrestrial stations can be optimized. The satellites are assumed
to follow Kepler orbits. The satellite visibility is defined as the long-term
mean in-view time from one or more terrestrial stations. The case of a single
satellite station is the basic building block for evaluating the regional
coverage of multi-satellite constellations.

It is found that the efficiency of optimizing the locations of terrestrial
stations is dependent on the specific satellite orbit and on the minimum
required elevation (MRE) of the satellite above the horizon. Selecting the
ground station locations becomes even more important when orthogonal time
sharing schemes are used with multi-satellite networks forming non-overlapping
visibility windows in time. The detailed description of Kepler orbits is
necessary to study satellite trajectories beyond simple circular orbits. Among
research problems involving satellite spatial geometries are satellite
communications. The orbits of satellite constellations affect the
characteristics of satellite communication channels, the terrestrial coverage
as well as topology and protocols of inter-satellite communication networks.
Even though commercial software for calculating precise satellite trajectories
exist, majority of researchers may prefer to implement Kepler orbits in their
simulations in order to avoid possibly large financial costs for acquiring
commercial software. It is of course possible to extend the model of Kepler
orbits with additional features to improve the modeling accuracy of satellite
trajectories. For example, the oblateness of the Earth induces drifts in the
Kepler orbits, and for lower orbits, the atmospheric drag cannot be neglected.

The rest of this paper is organized as follows. Section 2 presents a detailed
step-by-step model of Kepler orbits representing the satellite trajectory in 3D
space including the corresponding 2D trajectory on the Earth surface. The
satellite visibility is defined and analyzed in Section 3. Discussion in
Section 4 concludes the paper. In our mathematical notation, $\norm{\vv}$
denotes the Euclidean norm of vector $\vv$, $\hvv=\vv/\norm{\vv}$ is the unit
length vector, $\vv_1\times\vv_2$ and $\vv_1\cdot\vv_2$ are the cross product
and dot product of two vectors, respectively, and the components of 3D vectors
are denoted as $\vv=(v_x,v_y,v_z)$.

%%%%%
\section{Kepler orbits of LEO satellites}
%%%%%

In many communications scenarios, it is sufficient to model satellite movements
by assuming Kepler's two-body mechanics. The resulting Kepler orbit describes
the time dependent satellite trajectory in 3D space. Kepler orbit can be fully
defined by the two vectors: initial position $\rv(0)$ and initial speed
$\vv(0)$ at time $t=0$. These vectors represent a simple model of launching the
satellite into a given orbit. The position and speed vectors $\rv(t)$ and
$\vv(t)$, respectively, are normally expressed in the Earth-centered inertial
(ECI) coordinates having the origin at the center of mass of the Earth while
the axes do not rotate with the Earth \cite{wkp:ecicor}. The vectors $\rv(0)$
and $\vv(0)$ can be used to calculate the six basic elements of Kepler orbit
\cite{chobotov02}. The values of these elements fully define the orbit, so they
can be specified instead of the initial vectors. Specifically, semi-major axis
and eccentricity, respectively, describe the orbit size and shape. Inclination,
right ascension (or longitude) of the ascending node (RAAN) and argument of
perigee describe the orbit orientation in 3D space. Finally, true anomaly
determines the satellite position in its orbit. Thus, the space of all Kepler
orbits can be searched using either polar coordinates of vectors $\rv$ and
$\vv$, or using the six orbital elements.

Eccentricity $e=\norm{\ev}$ is the magnitude of eccentricity vector
\cite{wkp:excvec,wkp:orbexc,wkp:excanom},
\begin{equation*}
  \ev= \vv\times \hv/\mu-\hrv
\end{equation*}
where the angular momentum vector $\hv=\rv\times\vv$ \cite{wkp:angmom}, and
$\mu= G\cdot M_E= 398600.4$ $\units{km^3/s^2}$ is the standard gravitational
parameter \cite{wkp:standardgravpar}. The inclination angle $\iota$ of the
orbit is computed as, $\iota= \cos^{-1} (\hhv\cdot\kv_z)$ \cite{wkp:orbinc}
where the unit vector $\kv_z=(0,0,1)$. Using the vis-viva equation
\cite{wkp:visviva}, the semi-major axis $a$ is calculated as
\cite{wkp:semaxes},
\begin{equation*}
  a= \left(2/\norm{\rv} - \norm{\vv}^2/\mu\right)^{-1}.
\end{equation*}

The RAAN angle $\Omega$ is calculated by first defining the vector $\nv$
pointing towards the ascending node \cite{wkp:lonanode}, i.e., let
$\nv=\kv_z\times \hv$, and,
\begin{eqnarray*}
  \Omega=\left\{ \begin{array}{cc} \cos^{-1}(n_x/\norm{\nv}) & n_y\geq 0 \\
                   2\pi-\cos^{-1}(n_x/\norm{\nv}) & \mbox{otherwise}.
                 \end{array} \right.
\end{eqnarray*}

The argument of periapsis $\omega$ is the angle from the ascending node to its
periapsis measured in the direction of its motion \cite{wkp:argper}, i.e., it
is the angle between vectors $\nv$ and $\ev$ which can be calculated as,
\begin{equation*}
  \omega= \cos^{-1}(\hnv\cdot\hev).  
\end{equation*}
In order to determine the correct quadrant for $\omega$, we can also calculate,
\begin{equation*}
  \omega = \sin^{-1}\left(\frac{|\hnv\times \hev|}{|\hnv||\hev|}\right).
\end{equation*}

Finally, true anomaly $\nu$ is the angle between direction of periapsis and the
current satellite position \cite{wkp:trueanom}, i.e., it is the angle between
the vectors $\ev$ and $\rv$. This angle can be calculated as,
\begin{eqnarray*}
  \nu= \left\{\begin{array}{cc} \cos^{-1}\left(\frac{\ev}{e}\cdot\hrv\right) &
   \rv\cdot\vv\geq 0 \\ 
    2\pi-\cos^{-1}\left(\frac{\ev}{e}\cdot\hrv\right) &
   \mbox{otherwise}. \end{array} \right. 
\end{eqnarray*}

In addition to the six basic orbital elements
$\{e, \iota, a, \Omega, \omega, \nu\}$, it is useful to define the following
parameters describing the satellite trajectory. In particular, mean anomaly is
the angle the satellite would move along a hypothetical circular orbit having
the same speed and the same orbital period \cite{wkp:meananom}. It is
calculated from the eccentric anomaly $E$ which is related to the true anomaly
as,
\begin{equation*}
  \tan \frac{\nu}{2} = \sqrt{\frac{1+e}{1-e}}\,\tan \frac{E}{2},\mbox{ and }
  \sin E = \frac{\sqrt{1-e^2}\sin\nu}{1+e\,\cos\nu}.
\end{equation*}
The mean anomaly is then calculated as,
\begin{equation*}
  M=E-e\,\sin E.
\end{equation*}
Numerically, it is beneficial to calculate the mean anomaly $M_0$ at the
initial epoch (i.e., the reference time) $t_0$, and then update the mean
anomaly at time $t$ as \cite{wkp:meanmotion},
\begin{equation*}
  M(t)=M_0+n(t-t_0)  
\end{equation*}
where $n=\sqrt{\mu/a^3}$ is the satellite angular frequency (i.e., the number
of revolutions per unit of time). The new value of $E$ can be efficiently
calculated from the updated mean anomaly using, for example, the Newton-Raphson
iterations \cite{wkp:newrap}.

The instantaneous radius $r=\norm{\rv}$ of the satellite orbit is calculated
as, $r = a(1-e\cos E)$, where $r=R_E+h_s$, $R_E=6,378.1$ km is the Earth radius
\cite{wkp:earthrad}, and $h_s$ is the satellite altitude above the Earth
surface. The semi-minor axis of Kepler orbit is computed as, $b=a\sqrt{1-e^2}$.
The orbital period $T_0$ of the satellite is given by the Kepler's third law
\cite{wkp:keplaws}, i.e.,
\begin{equation*}
  T_0= \sqrt{\frac{a^3 4 pi^2}{\mu}}.  
\end{equation*}

Greenwich mean sidereal time (GMST) is used to measure longitude of the vernal
equinox (i.e., the angle from prime meridian), accounting for the number of
Earth rotations (measured in days) $J_{2000}$ since the 1st January 2000
\cite{wkp:sidertime}. Assuming the angular speed of Earth rotations,
$\omega_E=\num{7.292116e-5}$ $\units{rad/s}$ \cite{wkp:earthrot}, the sidereal
day duration of $86164$ seconds, and $G_0=1.727564$ being the Greenwich hour
angle (GHA) corresponding to the 1st second of year 2000, we have,
\begin{equation*}
  \GMST_0 = G_0+86164\,\omega_E(J_{2000}+1/2)\mod 2\pi.
\end{equation*}
The values of $\GMST$ are then simply updated as,
\begin{equation*}
  \GMST(t)= \GMST_0+ (t-t_0)2\pi/86164.
\end{equation*}

Finally, the instantaneous satellite position is expressed in the
Earth-centered Earth-fixed (ECEF) coordinates \cite{wkp:ecefcor}. The ECEF
coordinates have the origin at the Earth center of mass, and they rotate with
the Earth. The $x$-axis is passing through the vernal equinox (one of the two
crossings of the ecliptic and the celestial equators), the $z$-axis is aligned
with the Earth rotation axis, and the $y$-axis is perpendicular to the other
two. Thus, after updating true anomaly $\nu$, and the current radius $r$, the
current ECEF coordinates of the satellite are expressed in polar coordinates
as,
\begin{eqnarray*}
  X_\ecef &=&  r \left(\cos(\omega + \nu)\cos(\Omega) -
             \sin(\omega+\nu)\sin(\Omega)\cos(\iota)\right)\\
  Y_\ecef &=&  r \left(\cos(\omega + \nu)\sin(\Omega) +
             \sin(\omega+\nu)\cos(\Omega)\cos(\iota)\right)\\
  Z_\ecef &=& r \sin(\omega + \nu)\sin(\iota).
\end{eqnarray*}

%%%%%
\section{Satellite visibility}
%%%%%

The ground track of the satellite is obtained by converting the ECEF
coordinates to latitude $\phi_s$, longitude $\lambda_s$, and altitude $h_s$
using vectors,
\begin{eqnarray*}
  \gv &=& (\cos(\GMST),\sin(\GMST),0) \\
  \sv_{12} &=& (X_\ecef,Y_\ecef,0) \\
  \cv &=& \gv\times\sv_{12}=(c_x,0,0).
\end{eqnarray*}
We then obtain the ground track coordinates,
\begin{eqnarray*}
  \phi_s &=& \tan^{-1}\left( \frac{Z_\ecef}{X_\ecef^2+Y_\ecef^2}\right)\\
  \lambda_s &=& \sign{c_z}\, \mathrm{atan2}(\norm{\cv},\gv\cdot\sv_{12})\mod
                2\pi 
\end{eqnarray*}
where $\mathrm{atan2}$ is the 4-quadrant arc-tangent function, $\sign{\cdot}$
is the sign function, and the altitude $h_s=\norm{\rv}-R_E$. Our goal is to
assess the satellite visibility from a ground station at specific location on
the Earth surface at latitude $\phi_p$, longitude $\lambda_p$, and altitude
$h_p=0$. The Earth rotation is accounted for by first computing the RAAN value,
\begin{equation*}
  \Omega_p = \mathrm{atan2}(\sin(\lambda_p+\GMST),\cos(\lambda_p+\GMST)).  
\end{equation*}
The ECI coordinates of the selected ground station location are given by the
polar to Cartesian transformation,
\begin{eqnarray*}
  X_p &=& (h_p+R_E)\cos(\Omega_p)\cos(\phi_p)\\
  Y_p &=& (h_p+R_E)\sin(\Omega_p)\cos(\phi_p)\\
  Z_p &=& (h_p+R_E)\sin(\phi_p).
\end{eqnarray*}

To decide on satellite visibility from a ground station, define vectors
$\sv=(X_\ecef,Y_\ecef,Z_\ecef)$ of the current satellite orbital position, and
$\pv=(X_p,Y_p,Z_p)$ of the observation point on the Earth surface. Assuming a
2D plane touching the Earth at the point $\pv$, the satellite is visible as
long as it is above the horizon represented by this plane. However, in
realistic scenarios, local obstacles, satellite altitude and elevation as well
as antenna radiation and reception patterns dictate the actual MRE of the
satellite above the horizon in order it becomes visible at the ground station.
Denoting such MRE value as $\mre$, the satellite visibility condition can be
expressed mathematically using the cosine law, i.e.,
\begin{equation}\label{eq:10}
  C_1:\ \frac{(\sv-\pv)}{d_{sp}}\cdot \frac{\pv}{R_E}\geq
  \cos\left(\frac{\pi}{2}-\mre\right)
\end{equation}
where $d_{sp}=\norm{\sv-\pv}$ denotes the distance between the satellite and
the observation point, and we assume $\norm{\pv}=R_E$. Consequently, by
increasing the value $\mre$, the satellite visibility time window during one
overpass of the ground station is reduced by a factor,
\begin{equation*}
  f(\mre)=1 - \frac{2}{\pi}\,\mre.
\end{equation*}

%%%%%
\subsection{Numerical analysis of LEO satellite visibility}
%%%%%

Using the expressions defining Kepler orbit given in Section 2, it is
straightforward to verify that the minimum satellite orbital speed to sustain
its orbit is increasing with decreasing altitude. For instance, assuming
circular orbit, the speeds greater than $\norm{\vv}\geq 7.1$ $\units{km/s}$ are
required for sustaining the orbital altitudes below $1530$ km. The orbits above
200 km up to about 2000 km above the Earth are classified as low Earth orbits.
The sample footprint and altitude of a LEO satellite are shown in \fref{fig1x}
assuming the initial position and speed vectors $\rv=(700+R_E,0,0)$ [km] and
$\vv=7.5\times(0,\sqrt{2}/2, \sqrt{2}/2)$ [km/s]. For these initial vectors,
the orbital period is $T_0=98.59$ min, i.e., about $14.56$ rotations in a day
(i.e., in 86164 sec), so the satellite trajectory is not ground periodic (also
referred to as Earth-repeat, or Earth synchronous \cite{itu97}). The Earth
periodic orbit has to account for the Earth oblateness effect. For other
orbits, the number of revolutions for the satellite to pass again over the same
ground point is large (typically greater than $100$). We observe in
\fref{fig1x} that latitude trajectory as well as altitude changes can be well
approximated by a sine waveform (however, for other orbits, the sine
approximation may not be valid). Note also that, for Kepler orbits, the maximum
and the minimum attainable latitudes are equal. The chosen initial conditions
yield the elliptical orbit, so the satellite altitude appears periodic. The
longitude trajectory appears periodic, since it is considered modulo $2\pi$,
and one of its periods can be approximated by the exponential function in time
$t$ as,
\begin{equation*}
  \lambda_s \approx \pi\,\sign{t}\Big|\frac{t}{T_\lambda}\Big|^{\alpha}  
\end{equation*}
where $|\cdot|$ denotes the absolute value, $T_\lambda$ is the period, and
$0<\alpha<1$ is the fitting constant.

The attainable altitudes of the satellite can be adjusted by changes in the
speed magnitude $\norm{\vv}$. On the other hand, the initial speed direction
affects the attainable ground track latitudes. In particular, for Kepler orbit,
the highest latitude is equal to the inclination, and it is symmetric about the
equator. These dependencies are illustrated in \fref{fig2ed} where we assumed
the initial speed vector, $\vv=\norm{\vv}\times(0,v_y,v_z)$ where
$v_y=\cos\angle\vv$ and $v_z=\sin\angle\vv$. In this case, the inclination,
$\iota=\angle\vv$. Some theoretical calculations of the maximum attainable
latitudes are presented in \cite[p. 168 and 179]{capderou05}.

Given an observation point on the Earth surface, the periodicity of ground
satellite track such as the one shown in \fref{fig1x} makes the condition
\eref{eq:10} to be satisfied in bursts of time intervals. This is indicated by
periodically appearing clusters of $6$ red bars in the lower sub-figure of
\fref{fig5ed} assuming the satellite trajectory from \fref{fig1x}, and
$\mre=5^{\circ}$, $\phi_p=-5^{\circ}$, and $\lambda_p=15^{\circ}$. The bars
correspond to the in-view time instances at the location $(\phi_p,\lambda_p)$,
i.e., when the condition \eref{eq:10} is satisfied. The average in-view
duration is $13.1$ min (i.e., the average width of red bars) whereas the
average spacing between visibility windows within a cluster is $95.3$ min
(i.e., the average separation of red bars). The average spacing between
clusters is $1437.8$ min (i.e., the average separation of red bar clusters).

The sparsity of relatively short visibility windows can be exploited to
interleave additional non-overlapping visibility windows from other satellites.
The so-called train satellite constellation when satellites are following the
same orbital trajectory with some delay is attractive for its easy of
maintenance. The blue and green bars in \fref{fig5ed} are the visibility
windows of the other two satellites on the same orbit, but delayed by $58$ min
and $754$ min after the first satellite, respectively. The first and the second
satellites provides better satellites visibility within the visibility cluster
whereas the third satellite makes the visibility intervals to be more evenly
distributed over time by increasing frequency of the visibility clusters.

As a case study, in the sequel, we assume the region of the whole central
Africa spanning approximately latitudes $\phi_p\in(-15,5)$ and longitudes
$\lambda_p\in(10,40)$ to study the visibility of a single LEO satellite. The
MTSV $\eta_s$ is a function of location and time, i.e.,
$\eta_s=f(\phi_p,\lambda_p;t)$. The MTSV can be then defined as,
\begin{equation}
  \aveta(\phi_p,\lambda_p)= \lim_{T\to\infty}\frac{1}{T}\int_{-T/2}^{T/2}
  \eta_s(\phi_p,\lambda_p;t)\df t = \Prob{(\phi_p,\lambda_p):C_1}
\end{equation}
where
\begin{equation*}
  \eta_s(\phi_p,\lambda_p;t) = \left\{ \begin{array}{cc} 1 & C_1
   \mbox{ satisfied} \\ 0 & \mbox{otherwise}. \end{array}\right.
\end{equation*}
In other words, $\aveta(\phi_p,\lambda_p)$ is the probability that the
satellite visibility condition \eref{eq:10} is satisfied at the selected ground
station location $(\phi_p,\lambda_p)$.

The mean visibility is shown in the upper sub-figure of \fref{fig5ed} over the
24 hour observation window across the whole region considered assuming the same
satellite orbit as in \fref{fig1x} (i.e., about $15$ satellite rotations in
one day with inclination $\iota=15^{\circ}$), and the conservative MRE value
$\mre=20^{\circ}$. The difference between the locations with the smallest
visibility of just $2.5\%$ and those with the largest visibility of $3.75\%$ is
$3.7/2.5=1.50$ times. Thus, even though the mean visibility is only several
percent, the differences in satellite visibility among different locations even
within a relatively small region can reach $50\%$ which can translate to $50\%$
larger transmission capacity. Furthermore, multiple satellites spaced over the
same or different orbits can be used to increase the regional visibility. For
instance, assuming 3 constellations of 6 satellites each, the satellite
visibility windows are shown in the lower sub-figure of \fref{fig5ed} as
discussed above. More importantly, it should be noted that the differences
between the best and the poorest satellite visibility locations are multiplied
by the number of visibility windows, i.e., the number of satellites in the
constellation and the number of constellations considered. Therefore, the
optimization of ground station locations is more crucial in scenarios with
multi-satellite constellations in order to use the time limited bandwidth
within the satellite network efficiently.

It is useful to investigate the dependency of MTSV values on orbital
parameters. Here, we investigate the MTSV dependency on orbit inclination
$\iota$ and the visibility parameter $\mre$, respectively. We again assume the
satellite orbit corresponding to the ground trajectory shown in \fref{fig1x}.
The MTSV $\aveta$ values versus inclination $\iota$ for two different values of
$\mre=20^{\circ}$ and $\mre=0^{\circ}$ are shown in \fref{fig3ed}. The maximum,
minimum and the average values of $\aveta$ in \fref{fig3ed} are obtained over
the whole region as in \fref{fig5ed} with $1^{\circ}$ resolution in both
latitude and longitude. The maximum MTSV monotonically decreases with
inclination. However, the minimum and average MTSV are first increasing before
they monotonically start to decrease with increasing inclination. For smaller
values of inclination, the differences among locations with the best and the
poorest satellite visibility are largest, so it becomes more important to
optimize the locations of satellite ground stations. Furthermore, the satellite
visibility decreases with $\mre$ as expected.

The MTSV versus $\mre$ is shown in \fref{fig4ed}. These curves complement the
results shown in \fref{fig3ed}. However, now, all MTSV statistics monotonically
decrease with $\mre$. The differences among the locations are larger for
smaller values of $\mre$. The corresponding empirically computed probability
that the average MTSV value over the whole region of interest is larger than
the visibility threshold $\etathr$ is plotted in the lower sub-figure of
\fref{fig4ed}. We observe that MTSV of at least $0.75\%$ is guaranteed at all
locations within the region considered.

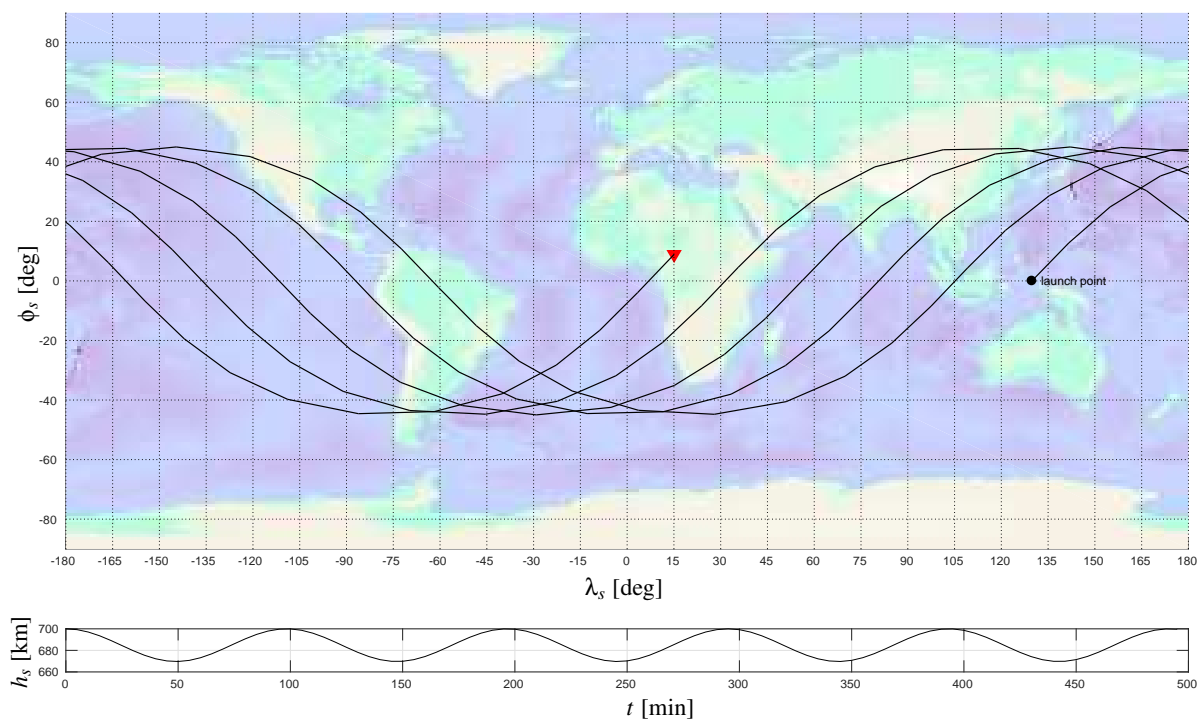
\xinsfig{-1.8}{scale=0.45}{fig1x}{The ground track of a LEO satellite (latitude
  vs. longitude) and the altitude for $5$ revolutions about the Earth.}{%
  \psfrag{yy1}[cc][][0.9][0]{$\phi_s$ [deg]}
  \psfrag{xx1}[cc][][0.9][0]{$\lambda_s$ [deg]}
  \psfrag{yy2}[cc][][0.9][0]{$h_s$ [km]}
  \psfrag{xx2}[cc][][0.9][0]{\quad\qquad$t$ [min]}}
    
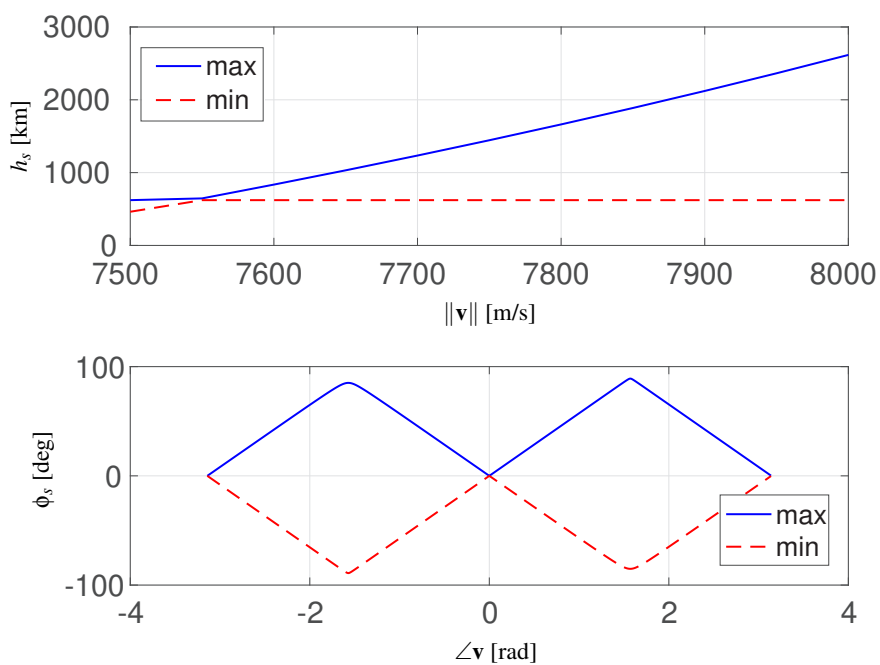
\xinsfig{0}{scale=0.75}{fig2ed}{(\textbf{upper sub-figure}) The maximum and
  minimum orbital altitudes vs. the initial speed magnitude for
  $\angle\vv=45^{\circ}$. (\textbf{lower sub-figure}) The maximum and minimum
  attainable latitudes vs. the initial speed direction for $\norm{\vv}=7.5$
  km/s.}{ \psfrag{yy1}[cc][][0.9][0]{$h_s$ [km]}
  \psfrag{xx1}[cc][][0.9][0]{$\norm{\vv}$ [m/s]}
  \psfrag{yy2}[cc][][0.9][0]{$\phi_s$ [deg]}
  \psfrag{xx2}[cc][][0.9][0]{$\angle \vv$ [rad]}}

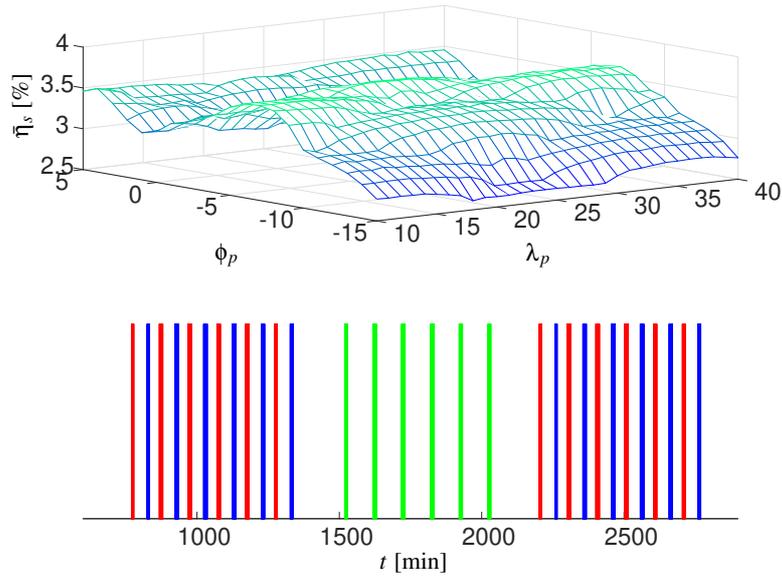
\xinsfig{0}{scale=0.75}{fig5ed}{The satellite follows the Kepler orbit with
  inclination $\iota=15^{\circ}$ and the MRE $\mre=20^{\circ}$. (\textbf{upper
    sub-figure}) The MTSV values accumulated over a $24$ hour observation
  window within the rectangular region with coordinates $(\phi_p,\lambda_p)$.
  (\textbf{lower sub-figure}) The satellite visibility time windows for 3
  satellite constellations of 6 satellites each denoted by blue, red and green
  bars, respectively, and the observer at location $(-5^{\circ},15^{\circ})$.}{
  \psfrag{yy1}[cc][][0.9][0]{$\phi_p$} \psfrag{xx1}[cc][][0.9][0]{$\lambda_p$}
  \psfrag{zz}[cc][][0.9][0]{$\aveta$ [\%]} \psfrag{xx2}[cc][][0.9][0]{$t$
    [min]}}

Finally, we note that the MTSV metric can be incorporated as a weighting factor
for other performance measures $\M_s$ such as the mean transmission distance
from the satellite or the capacity of satellite channel experienced at location
$(\phi_p,\lambda_p)$ on the ground. The mean value of such metric can be
calculated as the long-term time average, i.e.,
\begin{equation}
  \bar{M}_s(\phi_p,\lambda_p)=  \lim_{T\to\infty}\frac{1}{T}\int_{-T/2}^{T/2}
  \eta_s(\phi_p,\lambda_p;t) \,M_s(\phi_p,\lambda_p;t) \df t
\end{equation}
where the satellite visibility $\eta_s(\phi_p,\lambda_p;t)$ is used as a
weighting factor (or indicator, as in our definition, it is either equal to 1
or 0) of the metric of interest. Consequently, we can expect similar dependency
of the mean metric $\bar{M}_s$ on the orbital parameters as in case of the
MTSV.

%%%%%
\section{Discussion}
%%%%%

We investigated visibility of a single LEO satellite over a large geographical
area. The ground trajectory of the LEO satellite is modeled as the projection
of ideal Kepler orbit onto the Earth surface while neglecting the real world
orbital perturbations. Optimizing the satellite constellations or orbits to
maximize the coverage in a given region would not be economically viable unless
the region is sufficiently large and has high population density (for instance,
India and China may launch LEO satellites to provide the viable satellite
Internet access in their countries). However, the satellite networks are
normally assumed for more global coverage, and the end users can only optimize
locations of the ground stations. The satellite visibility in our paper is
measured using MTSV values. Other similar satellite visibility metrics can be
defined which may be more suitable in different scenarios. The satellite
visibility can appear as a weighting factor, for example, in evaluating the
area capacity of satellite links in a given region. Our numerical experiments
show that the long-term average values of MTSV within the large region can
differ by as much as $50\%$. The MTSV values are dependent on the satellite
orbit and the chosen value of MRE. Furthermore, the differences in MTSV among
different locations within the region are increasing proportionally with the
number of satellites and the number of satellite constellations.

Relatively short visibility windows of the overflying satellites require that
the satellites are closely spaced on multiple orbits to create a continuous
coverage of the whole Earth surface as indicated in \fref{pict1}. In order to
manage traffic routing among the relatively large number of satellites, the
satellite networks can be organized hierarchically into several tiers deployed
at different orbital altitudes, so only the lowest tier being closest to the
Earth surface communicate with ground stations. The ground footprints of these
lowest tier satellites are partially overlapping which requires careful
management of transmission channels to avoid possible inference issues in these
overlapping areas. It may be useful to form separate transmission and reception
antenna beams at satellite stations to exploit a full capacity of the
interference-free non-overlapping regions. The number of satellites $\Nsat$
required for the continuous Earth coverage can be estimated as,
\begin{equation*}
  \Nsat \approx \frac{4\pi R_E^2}{\pi\Rfp^2}(1+\kappa) =
  (4+4\kappa)\left(\frac{R_E}{\Rfp}\right)^2
\end{equation*}
where $R_E/\Rfp$ is the Earth radius normalized by the radius of the satellite
footprint, and $\kappa$ is the overlapping factor expressed as the percentage
of single satellite footprint. Assuming $\Rfp=100$ km, and $\kappa=50\%$, we
get, $\Nsat\approx 24,400$ satellites. This is in agreement with the recent
reports by several satellite network providers to launch about $25,000$ small
LEO satellites over the next several years.

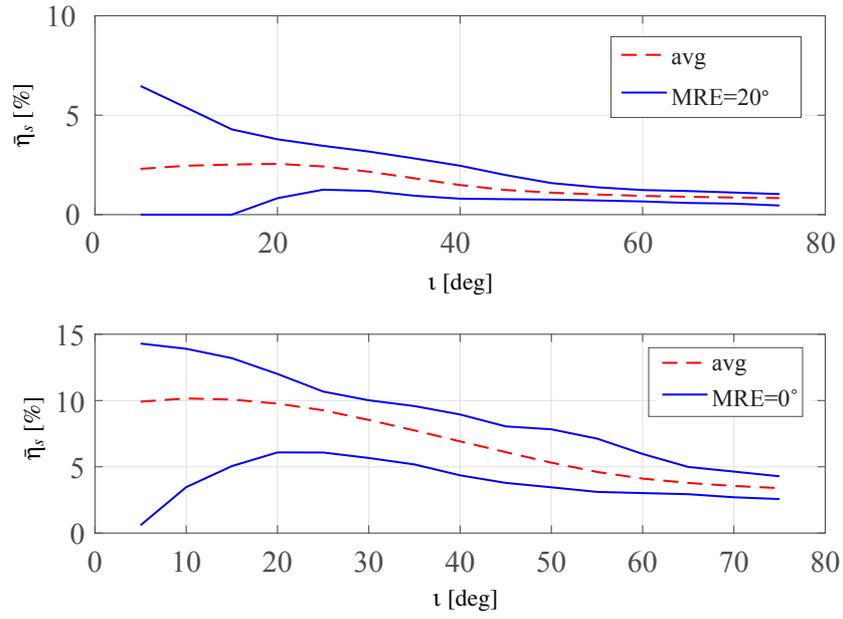
\xinsfig{0}{scale=0.75}{fig3ed}{The maximum, minimum and the average MTSV vs.
  orbit inclination. (\textbf{upper sub-figure}) The MRE value
  $\mre=20^{\circ}$. (\textbf{lower sub-figure}) The MRE value
  $\mre=0^{\circ}$.}{ \psfrag{yy1}[cc][][0.9][0]{$\aveta$ [\%]}
  \psfrag{xx1}[cc][][0.9][0]{$\iota$ [deg]} \psfrag{yy2}[cc][][0.9][0]{$\aveta$
    [\%]} \psfrag{xx2}[cc][][0.9][0]{$\iota$ [deg]}}

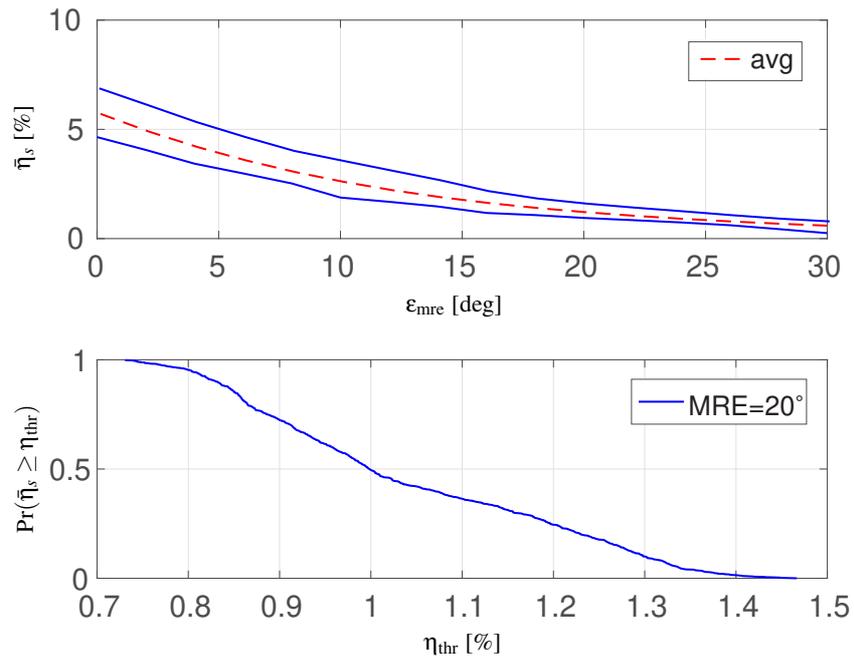
\xinsfig{0}{scale=0.75}{fig4ed}{(\textbf{upper sub-figure}) The maximum,
  minimum and the average MTSV vs. MRE for inclination $\iota=45^{\circ}$.
  (\textbf{lower sub-figure}) The probability that the average MTSV is above
  the threshold over the whole rectangular area considered assuming
  $\mre=20^{\circ}$.}{ \psfrag{yy1}[cc][][0.9][0]{$\aveta$ [\%]}
  \psfrag{xx1}[cc][][0.9][0]{$\mre$ [deg]}
  \psfrag{yy2}[cc][][0.9][0]{$\Prob{\aveta\geq \etathr}$}
  \psfrag{xx2}[cc][][0.9][0]{$\etathr$ [\%]}}

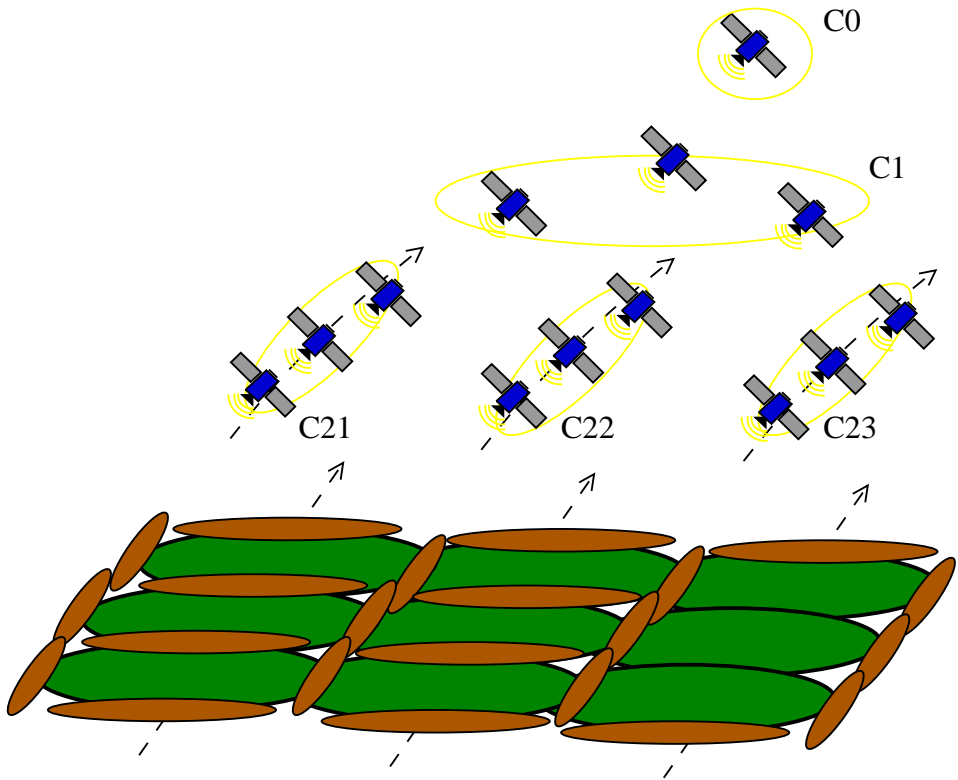
\insfig{0}{scale=1.5}{pict1}{Three tiers of satellite constellations denoted as
  C0, C1 and C2x, and the corresponding moving ground footprints with
  overlapping (brown color) and non-overlapping (green color) regions.}

Since architectures of existing cellular networks are well understood,
equivalent solutions can be considered also for communication networks
comprising the terrestrial and satellite segments. In particular, satellites
can act as base stations for the terrestrial users who are mobile with respect
to the satellites. However, as satellites are getting smaller to reduce the
infrastructure costs, such satellites have limited power supply and small
antennas, and cannot provide sufficiently strong signals to all ground users in
the area. Thus, small satellite designs constrain the number of ground users
being able to directly connect to the same satellite at the same time. These
issues can be overcome by aggregating traffic of terrestrial users using a
relatively small number of terminal stations on the ground. These ground
stations act as base stations and interconnect mobile users on the ground with
moving satellites. More importantly, the ground stations can exploit large
antenna arrays with beamforming to simultaneously track one or more satellites
in the sky, and they can implement other techniques developed previously for
cellular systems such as coordinated multi-point transmissions to improve the
reliability of communication links to satellites. Moreover, since terrestrial
transmissions are likely to be cheaper than the transmissions via satellite
relays, the key communication service provided by the satellite relays is to
interconnect otherwise isolated sub-networks, especially if such terrestrial
connections would be difficult to implement. \fref{pict2} shows the satellite
communication bridge connecting two mutually isolated autonomous systems
separated by the high mountains, the sea, or other such natural obstacles where
using satellites may be the only viable option. Future work may assume
constellation design for the ground stations as a counterpart of satellite
constellations, especially with large MRE values due to antenna radiation
patterns.

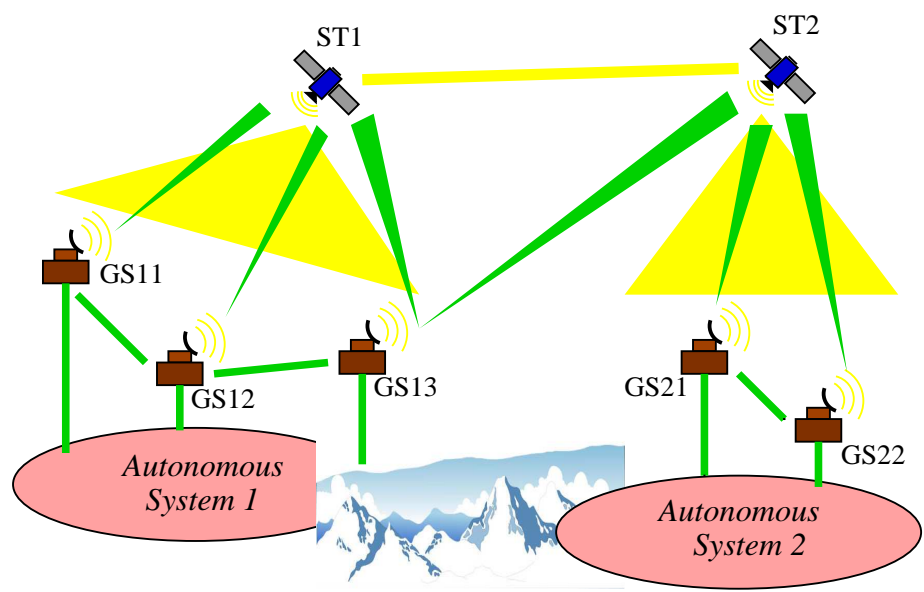
\insfig{0}{scale=1.5}{pict2}{The satellite communication bridge interconnecting
  two autonomous systems separated by the natural obstacle (e.g., mountains,
  sea), so the terrestrial connection between the ground stations (GS) in
  different autonomous systems would be too expensive or too difficult.}

%%%%%
\reftitle{References}
\bibliographystyle{mdpi}
\bibliography{refer}\bigskip

\begin{thebibliography}{-------}
\providecommand{\natexlab}[1]{#1}

\bibitem[Crisp \em{et~al.}(2015)Crisp, Smith, and Hollingsworth]{crisp2015}
Crisp, N.; Smith, K.; Hollingsworth, P.
\newblock Launch and deployment of distributed small satellite systems.
\newblock {\em Acta Astronautica} {\bf 2015}, {\em 114},~65--78.

\bibitem[Khan(2016)]{khan16}
Khan, F.
\newblock Multi-Comm-Core Architecture for Terabit-per-Second Wireless.
\newblock {\em IEEE Communications. Magazine} {\bf 2016}, {\em 54},~124--129.

\bibitem[Lee and Park(2019)]{lee2019}
Lee, K.H.; Park, K.Y.
\newblock Overall Design of Satellite Networks for Internet Services with {QoS}
  Support.
\newblock {\em Electronics} {\bf 2019}, {\em 8},~1--21.

\bibitem[Wang \em{et~al.}(2019)Wang, Wang, and Wang]{wang2019}
Wang, C.; Wang, H.; Wang, W.
\newblock A Two-Hops State-Aware Routing Strategy Based on Deep Reinforcement
  Learning for LEO Satellite Networks.
\newblock {\em Electronics} {\bf 2019}, {\em 8},~1--17.

\bibitem[Clifford \em{et~al.}(1994--2002)Clifford, Pontieu, and Hunt]{satobs}
Clifford, N.; Pontieu, B.D.; Hunt, J.
\newblock Visual Satellite Observer.
\newblock \url{http://www.satobs.org/}.

\bibitem[Zantou \em{et~al.}(2005)Zantou, Kherras, and Addaim]{zantou2005}
Zantou, E.B.; Kherras, A.; Addaim, A.
\newblock Orbit Calculation and Doppler Correction Algorithm in a LEO Satellite
  Small Ground Terminal.
\newblock  19th Annual AIAA/USU Conference on Small Satellites,  2005, pp.
  1--8.

\bibitem[Chobotov(2002)]{chobotov02}
Chobotov, V.A.
\newblock {\em Orbital Mechanics}, 3rd ed. ed.; AIAA,  2002.

\bibitem[Vallado and McClain(1997)]{vallado1997}
Vallado, D.A.; McClain, W.D.
\newblock {\em Fundamentals and Astrodynamics and Applications}, 1st ed. ed.;
  McGraw Hill,  1997.

\bibitem[Bowman and Storz(2003)]{bowman2003}
Bowman, B.R.; Storz, M.F.
\newblock High accuracy satellite drag model {(HASDM)} review.
\newblock {\em American Astronautical Society} {\bf 2003}, {\em 03-625},~1--10.

\bibitem[Gergely(2015)]{gergely2015}
Gergely, L.A.
\newblock Celestial mechanics: {T}he perturbed {K}epler problem.
\newblock Technical report, Univ. of Szeged,  2015.

\bibitem[Capderou(2005)]{capderou05}
Capderou, M.
\newblock {\em Satellites Orbits and Missions}; Springer,  2005.

\bibitem[Ballard(1980)]{ballard80}
Ballard, A.
\newblock Rosette Constellations of {E}arth Satellites.
\newblock {\em IEEE Transactions on Aerospace and Electronic Systems} {\bf
  1980}, {\em AES-16},~656--673.

\bibitem[Danesfahani and Pirhadi(2006)]{danesf2006}
Danesfahani, R.; Pirhadi, M.
\newblock A Novel Constellation of Satellites.
\newblock  ICICT,  2006, pp. 2491--2495.

\bibitem[Werner \em{et~al.}(1995)Werner, Jahn, Lutz, and
  B\"ottcher]{werner1995}
Werner, M.; Jahn, A.; Lutz, E.; B\"ottcher, A.
\newblock Analysis of System Parameters for LEO/ICO-Satellite Communication
  Networks.
\newblock {\em IEEE Journal on Selected Areas in Communications} {\bf 1995},
  {\em 13},~371--381.

\bibitem[Chen \em{et~al.}(2017)Chen, Dai, Reinelt, and Wang]{chen17}
Chen, X.; Dai, G.; Reinelt, G.; Wang, M.
\newblock A Semi-Analytical Method for Periodic {E}arth Coverage Satellites
  Optimization.
\newblock {\em IEEE Communications Letters} {\bf 2017}, {\em 22},~534--537.

\bibitem[{Rec. ITU-R S.1257}(1997)]{itu97}
{Rec. ITU-R S.1257}.
\newblock Analytical Method To Calculate Visibility Statistics For
  Non-Geostationary Satellite Orbit Satellites As Seen From A Point On The
  {E}arth's Surface.
\newblock Technical report, ITU,  1997.

\bibitem[Lyras \em{et~al.}(2018)Lyras, Efrem, Kourogiorgas, and
  Panagopoulos]{lyras2018}
Lyras, N.K.; Efrem, C.N.; Kourogiorgas, C.I.; Panagopoulos, A.D.
\newblock Optimum Monthly-Based Selection of Ground Stations for Optical
  Satellite Networks.
\newblock {\em IEEE Communications Letters} {\bf 2018}, {\em 22},~1192--1195.

\bibitem[Ali \em{et~al.}(1999)Ali, Al-Dhahir, and Hershey]{ali99}
Ali, I.; Al-Dhahir, N.; Hershey, J.E.
\newblock Predicting the Visibility of {LEO} Satellites.
\newblock {\em IEEE Transactions on Aerospace and Electronic Systems} {\bf
  1999}, {\em 35},~1183--1190.

\bibitem[Ali \em{et~al.}(1998)Ali, Al-Dhahir, and Hershey]{ali1998}
Ali, I.; Al-Dhahir, N.; Hershey, J.E.
\newblock Doppler Characterization for LEO Satellites.
\newblock {\em IEEE Transactions on Communications} {\bf 1998}, {\em
  46},~309--313.

\bibitem[Pratt \em{et~al.}(1999)Pratt, Raines, {Fossa Jr.}, and
  Temple]{pratt99}
Pratt, S.R.; Raines, R.A.; {Fossa Jr.}, C.E.; Temple, M.A.
\newblock An Operational and Performance Overview of the {I}ridium Low {E}arth
  Orbit Satellite Systems.
\newblock {\em IEEE Communication Surveys and Tutorials} {\bf 1999}, {\em
  2},~2--10.

\bibitem[Cakaj \em{et~al.}(2014)Cakaj, Kamo, Lala, and Rakipi]{cakaj14}
Cakaj, S.; Kamo, B.; Lala, A.; Rakipi, A.
\newblock The Coverage Analysis for Low {E}arth Orbiting Satellites at Low
  Elevation.
\newblock {\em International Journal of Advanced Computer Science and
  Applications} {\bf 2014}, {\em 5},~6--10.

\bibitem[Seyedi and Safavi(2012)]{seyedi12}
Seyedi, Y.; Safavi, S.M.
\newblock On the Analysis of Random Coverage Time in Mobile {LEO} Satellite
  Communications.
\newblock {\em IEEE Communications Letters} {\bf 2012}, {\em 16},~612--615.

\bibitem[Hossain \em{et~al.}(2017)Hossain, Hassan, Atiquzzaman, and
  Ivancic]{hossain2017}
Hossain, M.S.; Hassan, S.S.; Atiquzzaman, M.; Ivancic, W.
\newblock Survivability and scalability of space networks: a survey.
\newblock {\em Telecommunication Systems} {\bf 2017}.
\newblock
  doi:{\changeurlcolor{black}\href{https://doi.org/10.1007/s11235-017-0396-y}{\detokenize{10.1007/s11235-017-0396-y}}}.

\bibitem[Lutz \em{et~al.}(1991)Lutz, Cygan, Dippold, Dolainsky, and
  Papke]{lutz1991}
Lutz, E.; Cygan, D.; Dippold, M.; Dolainsky, F.; Papke, W.
\newblock The Land Mobile Satellite Communication Channel-Recording,
  Statistics, and Channel Model.
\newblock {\em IEEE Transactions on Vehicular Technology} {\bf 1991}, {\em
  40},~375--386.

\bibitem[Abdi \em{et~al.}(2003)Abdi, Lau, Alouini, and Kaveh]{abdi2003}
Abdi, A.; Lau, W.C.; Alouini, M.S.; Kaveh, M.
\newblock A New Simple Model for Land Mobile Satellite Channels: {F}irst- and
  Second-Order Statistics.
\newblock {\em IEEE Transactions on Communications} {\bf 2003}, {\em
  2},~519--528.

\end{thebibliography}


\begin{thebibliography}{10}
  \raggedright
  
\bibitem{wkp:angmom} Wikipedia contributors, ``Angular momentum,''
  \textit{Wikipedia}, \textit{The Free Encyclopedia},
  \url{https://en.wikipedia.org/w/index.php?title=Angular_momentum} (accessed
  September 17, 2019).

\bibitem{wkp:argper} Wikipedia contributors, ``Argument of periapsis,''
  \textit{Wikipedia}, \textit{The Free Encyclopedia},
  \url{https://en.wikipedia.org/w/index.php?title=Argument_of_periapsis}
  (accessed September 17, 2019).
  
\bibitem{wkp:ecefcor} Wikipedia contributors, ``Earth-centered, Earth-fixed,''
  \textit{Wikipedia}, \textit{The Free Encyclopedia},
  \url{https://en.wikipedia.org/w/index.php?title=ECEF} (accessed September 17,
  2019).

\bibitem{wkp:ecicor} Wikipedia contributors, ``Earth-centered inertial,''
  \textit{Wikipedia}, \textit{The Free Encyclopedia},
  \url{https://en.wikipedia.org/w/index.php?title=Earth-centered_inertial}
  (accessed September 17, 2019).

\bibitem{wkp:earthrad} Wikipedia contributors, ``Earth radius,''
  \textit{Wikipedia}, \textit{The Free Encyclopedia},
  \url{https://en.wikipedia.org/w/index.php?title=Earth_radius} (accessed
  September 17, 2019).
  
\bibitem{wkp:earthrot} Wikipedia contributors, ``Earth's rotation,''
  \textit{Wikipedia}, \textit{The Free Encyclopedia},
  \url{https://en.wikipedia.org/w/index.php?title=Earth\%27s_rotation}
  (accessed September 17, 2019).
    
\bibitem{wkp:excanom} Wikipedia contributors, ``Eccentric anomaly,''
  \textit{Wikipedia}, \textit{The Free Encyclopedia},
  \url{https://en.wikipedia.org/w/index.php?title=Eccentric_anomaly} (accessed
  September 17, 2019).

\bibitem{wkp:excvec} Wikipedia contributors, ``Eccentricity vector,''
  \textit{Wikipedia}, \textit{The Free Encyclopedia},
  \url{https://en.wikipedia.org/w/index.php?title=Eccentricity_vector}
  (accessed September 17, 2019).

\bibitem{wkp:keplaws} Wikipedia contributors, ``Kepler's laws of planetary
  motion,'' \textit{Wikipedia}, \textit{The Free Encyclopedia},
  \url{https://en.wikipedia.org/w/index.php?title=Kepler\%27s_laws_of_planetary_motion}
  (accessed September 17, 2019).

\bibitem{wkp:lonanode} Wikipedia contributors, ``Longitude of the ascending
  node,'' \textit{Wikipedia}, \textit{The Free Encyclopedia},
  \url{https://en.wikipedia.org/w/index.php?title=Longitude_of_the_ascending_node}
  (accessed September 17, 2019).

\bibitem{wkp:meananom} Wikipedia contributors, ``Mean anomaly,''
  \textit{Wikipedia}, \textit{The Free Encyclopedia},
  \url{https://en.wikipedia.org/w/index.php?title=Mean_anomaly} (accessed
  September 17, 2019).

\bibitem{wkp:meanmotion} Wikipedia contributors, ``Mean motion,'' \textit{Wikipedia},
  \textit{The Free Encyclopedia},
  \url{https://en.wikipedia.org/w/index.php?title=Mean_motion} (accessed
  September 17, 2019).  

\bibitem{wkp:newrap} Wikipedia contributors, ``Newton's method,''
  \textit{Wikipedia}, \textit{The Free Encyclopedia},
  \url{https://en.wikipedia.org/w/index.php?title=Newton\%27s_method} (accessed
  September 17, 2019).

\bibitem{wkp:orbexc} Wikipedia contributors, ``Orbital eccentricity,''
  \textit{Wikipedia}, \textit{The Free Encyclopedia},
  \url{https://en.wikipedia.org/w/index.php?title=Orbital_eccentricity}
  (accessed September 17, 2019).
  
\bibitem{wkp:orbinc} Wikipedia contributors, ``Orbital inclination,''
  \textit{Wikipedia}, \textit{The Free Encyclopedia},
  \url{https://en.wikipedia.org/w/index.php?title=Orbital_inclination}
  (accessed September 17, 2019).

\bibitem{wkp:semaxes} Wikipedia contributors, ``Semi-major and semi-minor
  axes,'' \textit{Wikipedia}, \textit{The Free Encyclopedia},
  \url{https://en.wikipedia.org/w/index.php?title=Semi-major_and_semi-minor_axes}
  (accessed September 17, 2019).

\bibitem{wkp:sidertime} Wikipedia contributors, ``Sidereal time,''
  \textit{Wikipedia}, \textit{The Free Encyclopedia},
  \url{https://en.wikipedia.org/w/index.php?title=Sidereal_time} (accessed
  September 17, 2019).

\bibitem{wkp:standardgravpar} Wikipedia contributors, ``Standard gravitational
  parameter,'' \textit{Wikipedia}, \textit{The Free Encyclopedia},
  \url{https://en.wikipedia.org/w/index.php?title=Standard_gravitational_parameter}
  (accessed September 17, 2019).

\bibitem{wkp:trueanom} Wikipedia contributors, ``True anomaly,''
  \textit{Wikipedia}, \textit{The Free Encyclopedia},
  \url{https://en.wikipedia.org/w/index.php?title=True_anomaly} (accessed
  September 17, 2019).

\bibitem{wkp:visviva} Wikipedia contributors, ``Vis-viva equation,''
  \textit{Wikipedia}, \textit{The Free Encyclopedia},
  \url{https://en.wikipedia.org/w/index.php?title=Vis-viva_equation} (accessed
  September 17, 2019).

\end{thebibliography}

%%%%%
It should be noted that the following resources on Wikipedia are summarized
here mainly for the convenience. For more rigorous understanding of orbital
mechanics, the interested reader is referred to classical textbooks on this
topic, for example, \cite{chobotov02} and \cite{vallado1997} cited above.

\reftitle{Wikipedia references}
% Chicago style

\end{document}